\documentclass[aps,prl,twocolumn,groupedaddress,showpacs]{revtex4}
\usepackage{graphicx}% Include figure files
\usepackage{dcolumn}% Align table columns on decimal point
\usepackage{bm}% bold math
\begin{document}

\title{Can a wormhole be interpreted as an EPR pair?}

\author{Hrvoje Nikoli\'c}
\affiliation{Theoretical Physics Division, Rudjer Bo\v{s}kovi\'{c}
Institute,
P.O.B. 180, HR-10002 Zagreb, Croatia.}
\email{hrvoje@thphys.irb.hr}

\date{\today}

\begin{abstract}
Recently, Maldacena and Susskind arXiv:1306.0533 and Jensen and Karch arXiv:1307.1132
argued that a wormhole can be interpreted as an EPR pair.
We point out that a convincing justification of such an interpretation
would require a quantitative evidence that correlations between two ends of the wormhole 
are equal to those between the members of the EPR pair.
As long as the existing results do not contain
such evidence, the interpretation of wormhole as an EPR pair does not seem justified.
\end{abstract}


\maketitle

Recently, Maldacena and Susskind \cite{MS} conjectured that a wormhole can be interpreted
as an EPR pair. Inspired by this conjecture, Jensen and Karch \cite{JK} attempted
to make the conjecture more precise, by arguing that the holographic dual of an EPR pair has a wormhole.
In this brief comment we argue that the results presented in those two papers are still very far from
presenting convincing evidence that a wormhole can be interpreted as an EPR pair.

The distinguished feature of an EPR pair is the existence of highly nontrivial correlations between two
members of the pair. In particular, the EPR correlations violate Bell inequalities \cite{bell}.
Unfortunately, no such nontrivial correlations have been calculated in \cite{MS} and \cite{JK}. 

In \cite{JK}, it has been demonstrated that entanglement entropy associated with one member of 
the EPR pair coincides with entropy of the corresponding end of the wormhole.
Even though this result is interesting and somewhat surprising,
the entanglement entropy {\it per se} is a single number 
which does not contain much information about
the details of correlations between two subsystems. Two bipartite quantum systems may 
be characterized by the same entanglement entropy, and yet obey very different correlations between 
their respective subsystems. For example, the bipartite entangled states $|+\rangle$ and $|-\rangle$, 
defined as
\begin{equation}
 |\pm\rangle = |\psi_1\rangle |\psi_2\rangle \pm |\psi_2\rangle |\psi_1\rangle ,
\end{equation}
lead to the same entanglement entropy, and yet to
different correlations. In \cite{MS}, $|+\rangle$ and $|-\rangle$ are interpreted as two 
qualitatively different wormholes, but a quantitative 
formulation of the correspondence based on entanglement entropy \cite{JK} 
cannot make such a distinction.

Moreover, entanglement entropy is a property of a {\em reduced} density matrix, associated
with one of the subsystems. Such a reduced density matrix describes what can be said about 
this subsystem if the other subsystem is not measured at all. By contrast, correlations describe 
the relations between measurements on {\em both} subsystems. 

Just as a precise formulation of AdS/CFT correspondence requires a match between 
all correlation functions of the two theories \cite{witten}, a similar precise formulation
in terms of correlations should be required for the conjectured relation between 
wormholes and EPR pairs. Without any quantitative evidence for the match of correlations 
it is difficult to take the conjecture seriously. 

If such a required match between the correlations would be established in a future work, 
that would be truly surprising; arguably even more surprising than the match between 
the correlation functions in AdS/CFT \cite{witten}.
But as long as the existing results in \cite{MS} and \cite{JK} do not contain
any direct evidence for such a match in terms of correlations, 
the conjectured interpretation of wormhole as an EPR pair does not seem sufficiently 
justified.

This work was supported by the Ministry of Science of the
Republic of Croatia under Contract No.~098-0982930-2864.

\end{document}